\documentclass[10pt,prd,twocolumn]{revtex4}
\usepackage{latexsym}
\usepackage{epsfig}
\usepackage{amssymb}
\usepackage{amsthm}
\usepackage{amsmath}
\usepackage{amssymb,amsfonts}
\usepackage{verbatim}

\newcommand{\be}{\begin{equation}}
\newcommand{\ee}{\end{equation}}
\newcommand{\bea}{\begin{eqnarray}}
\newcommand{\eea}{\end{eqnarray}}
\newcommand{\nn}{\nonumber}

\DeclareMathOperator{\tr}{\textrm{tr}}
\DeclareMathOperator{\Tr}{\textrm{Tr}}

\DeclareMathOperator{\MA}{\mathcal{A}}

\DeclareMathOperator{\MD}{\mathcal{D}}

\DeclareMathOperator{\ML}{\mathcal{L}}
\DeclareMathOperator{\MM}{\mathcal{M}}
\DeclareMathOperator{\MN}{\mathcal{N}}

\DeclareMathOperator{\MP}{\mathcal{P}}
\DeclareMathOperator{\MQ}{\mathcal{Q}}
\DeclareMathOperator{\MR}{\mathcal{R}}

\DeclareMathOperator{\MBFS}{\mathbf{S}}

\DeclareMathOperator{\MBBR}{\mathbb{R}}
\DeclareMathOperator{\MBBZ}{\mathbb{Z}}

\begin{document}

%PRD Style
\title{A New Anomaly-Free Gauged Supergravity in Six Dimensions}%
\author{Spyros D. Avramis}%
\email{avramis@cern.ch}
\author{Alex Kehagias}%
\email{kehagias@cern.ch} \affiliation{Department of Physics,
National Technical University of Athens, GR-15773, Zografou, Athens,
Greece}
\author{S. Randjbar-Daemi}%
\email{seif@ictp.trieste.it} \affiliation{International Center for
Theoretical Physics, 34100, Trieste, Italy}

\hskip 12cm
\begin{flushright}
\end{flushright}
\hskip 4cm
\begin{flushright}
hep-th/0504033\\ IC/2005/13
\end{flushright}

\date{\today}%
\begin{abstract}
We present a new anomaly-free gauged $\MN=1$ supergravity model in
six dimensions. The gauge group is $E_7 \times G_2 \times U(1)_R$,
with all hyperinos transforming in the product representation
$(\mathbf{56},\mathbf{14})$. The theory admits monopole
compactifications to $\MBBR^4 \times \MBFS^2$, leading to $D=4$
effective theories with broken supersymmetry and massless
fermions.

\end{abstract}
\maketitle

\section{Introduction}
\label{sec-1}

Minimal supergravity theories in six dimensions have a remarkably
rich structure and have attracted much interest over the years.
Some of the reasons motivating the study of such theories are
their connection to superstring vacua, their relation to $\MN=2$
theories in $D=4$ and the framework they provide for cosmological
investigations. Among the most interesting models of this type are
$D=6$ gauged supergravity models which have the important property
that they spontaneously compactify on lower-dimensional spaces. A
prototype for such compactifications is employed in the
Salam-Sezgin model \cite{Salam:1984cj}, a $D=6$ supersymmetric
Einstein-Maxwell theory. This theory admits an $\MBBR^4 \times
\MBFS^2$ solution that preserves half the supersymmetries,
obtained through a magnetic monopole background residing on
$\MBFS^2$ and, as has recently been shown \cite{GGP}, it is the
unique maximally-symmetric solution in this class of models.

One further interesting  aspect of the monopole compactification
in the supersymmetric models is the possibility of making all the
$U(1)$ factors in the gauge group massive. In the
non-supersymmetric Einstein-Maxwell theory in six dimensions, the
monopole compactification gives rise to an effective $D=4$ chiral
gauge theory with gauge group $SU(2)_{KK}\times U(1)$. The
chirality of the $D=4$ effective theory is due to the $U(1)$
factor which remains massless and has complex representations. In
the supersymmetric generalizations, such as the Salam-Sezgin model
or the non-Abelian theories discussed in this paper, the vector
potentials associated to the $U(1)$ factors of magnetic monopoles
acquire a mass due to their Chern-Simons coupling to the
second-rank antisymmetric potentials. This coupling is an
essential ingredient of all such models.

In addition to the nonzero mass for the $U(1)$ gauge potentials in
the monopole directions, most $D=6$ supergravity theories,
including the Salam-Sezgin model, suffer from the breakdown of
local symmetries due to the presence of gravitational, gauge and
mixed anomalies \cite{Alvarez-Gaume:1983ig} which render such
theories inconsistent at the quantum level. In fact, anomaly
cancellation has turned out to be a crucial guiding principle for
the identification of consistent $D=6$ theories for the same
reason as in the $D=10$ case. Although the $D=6$ anomaly
cancellation conditions are weaker than those in $D=10$, they are
still very stringent, especially in the case of gauged
supergravity theories.

Regarding Poincar\'e supergravities, there are numerous
anomaly-free theories in the literature. Most of these were found
by compactifying heterotic string theory on $K3$ using various
methods \cite{Green:1984bx} of embedding the $K3$ holonomy group
in the $SO(32)$ or $E_8 \times E_8$ gauge group. There are also
theories with enhanced symmetry originating from the Gepner points
of orbifold realizations of $K3$ \cite{Erler:1993zy} or from the
non-perturbative mechanism of small instantons
\cite{Witten:1995gx}, as well as theories found by solving the
anomaly cancellation conditions alone \cite{Schwarz:1995zw}. Such
theories can also be constructed as boundary theories on the
six-dimensional orbifold fixed points in seven dimensions
compactified on $\MBFS^1/\MBBZ_2$
\cite{{Gherghetta:2002xf},{Avramis:2004cn}}.

However, in the case of \emph{gauged} supergravities, there is
only one known non-trivial anomaly-free model, namely the $E_7
\times E_6 \times U(1)_R$ model of \cite{Randjbar-Daemi:1985wc}.
This model contains 456 hypermatter fields, identified as a
half-hypermultiplet in the pseudoreal $\mathbf{912}$ of $E_7$. The
theory satisfies a set of highly non-trivial anomaly constraints
which make it possible to completely cancel all anomalies by the
Green-Schwarz mechanism. Moreover, gauging of $U(1)_R$ gives rise
to a positive-definite potential which implies that one may turn
on a magnetic monopole background in a $U(1)$ subgroup of the
gauge group and compactify the theory on $\MBBR^4 \times \MBFS^2$.
In the particular case considered in \cite{Randjbar-Daemi:1985wc},
where the monopole is embedded in the ``hidden'' $E_6$, the
fermionic zero modes come exclusively from the $E_6$ gauginos. The
resulting $D=4$ theory has an $SO(10) \times U(1)_R$ gauge
symmetry. Unlike the Salam-Sezgin model, supersymmetry is
completely broken. For the minimal value of the monopole number,
which is required by the stability of the compactification, one
obtains two chiral families of $SO(10)$ in the 16-dimensional
spinor representation.

One very attractive property of the minimal gauged supergravities
in $D=6$ is that, unlike the superstring theories in $D=10$, they
do not admit the flat spaces as their most symmetric solutions. On
the other hand the $\MBBR^4\times \MBFS^2$ configuration is the
unique maximally-symmetric compactification \cite{GGP}.
Furthermore, the expectation values of all the scalars in the
model, with the exception of the dilaton, are uniquely determined
at the tree level. Hence there is only one modulus accompanying
these vacua. Like the $D=10$ supergravities, these models also
admit brane solutions of various dimensions \cite{{Guven:2003uw},
{Aghababaie:2003ar}, {Randjbar-Daemi:2004qr},
{Burgess:2004dh},{GGP}}. Because of the uniqueness and simplicity
of the $\MBBR^4\times \MBFS^2$ compactification as well as many
shared features with the $D=10$ heterotic theory it is useful to
construct more models of this type and study their low-energy
physics.

In this paper, we demonstrate the existence of a new gauged
anomaly-free $D=6$, $\MN=1$ model, besides the known one of
\cite{Randjbar-Daemi:1985wc}. The gauge group here is $E_7 \times
G_2 \times U(1)_R$ and so the theory contains $148$ vector
multiplets. Restricting to onnly one tensor multiplet,
cancellation of the irreducible gravitational anomaly demands then
$392$ hypermultiplets. These fit exactly into a
half-hypermultiplet of the pseudoreal product representation
$(\mathbf{56},\mathbf{14})$ of $E_7 \times G_2$. Again, it is
remarkable that all anomalies of the theory cancel in a
non-trivial way. Moreover, the model is also free of global
anomalies \cite{Witten:1982fp,Bergshoeff:1986hv} that could
potentially arise due to the presence of the $G_2$ factor.
Regarding the bosonic sector, the $1568$ real hyperscalars
parameterize the quaternionic manifold $Sp(392,1) / Sp(392) \times
Sp(1)$ and gauging of $U(1)_R$ contained in $Sp(1)$ and $E_7
\times G_2$ contained in $Sp(392)$ yields a positive-definite
potential allowing $\MBBR^4 \times \MBFS^2$ compactifications.
Unlike the $E_7 \times E_6 \times U(1)_R$ case, there are
fermionic zero modes coming from both the gauginos of the $E_7
\times G_2$ subgroup where the monopole is embedded \emph{and} the
hyperinos, since the latter transform non-trivially under both
$E_7$ and $G_2$ factors. These compactifications generate a rich
spectrum of chiral fermions in the $\mathbf{27}$'s  (or
$\mathbf{16}$'s ) of the unbroken $E_6$ (or $SO(10)$). However, as
we shall show, they are perturbatively unstable. To find an
anomaly-free model with a realistic $D=4$ fermion spectrum still
remains a challenging and unsolved problem.

This paper is organized as follows. In Section \ref{sec-2} we fix
our notation, we describe the basic aspects of gauged $D=6$
supergravity theories and we write the bosonic Lagrangian of our
model. In Section \ref{sec-3} we explicitly show that the theory
is free of anomalies. In Section \ref{sec-4} we discuss
compactification of the theory on $\MBBR^4 \times \MBFS^2$ and we
briefly consider various aspects of the effective $D=4$ theory. In
Section \ref{sec-5} we discuss the spectrum of $D=4$ chiral
fermions in detail and show the existence of the tachyonic mode in
the spectrum of most compactifications. Finally, in Section
\ref{sec-6}, we summarize and conclude.

\section{The Model}
\label{sec-2}

The building blocks of $D=6$, $\MN=1$ supergravity theories are the
massless representations of the minimal supersymmetry algebra which
is chiral and has $Sp(1)$ as its R-symmetry group. The field content
of these representations is summarized in the following multiplets
\bea \label{e-2-1} \text{Supergravity multiplet} \quad&:&\quad
( g_{MN} , B^-_{MN} , \psi^{i-}_M ),\nn\\
\text{Tensor multiplet} \quad&:&\quad
( B^+_{MN} , \phi , \chi^{i+} ), \nn\\
\text{Vector multiplet} \quad&:&\quad
( A_M, \lambda^{i-} ), \nn\\
\text{Hypermultiplet} \quad&:&\quad ( 4 \varphi , 2 \psi^{+} ), \eea
where, the $+$ ($-$) superscripts denote positive (negative)
chirality for the spinors and (anti-)self-duality for 2--forms and
the index $i=1,2$ takes values in the fundamental of $Sp(1)_R$.

A general $D=6$, $\MN=1$ supergravity theory coupled to matter is
constructed by combining one supergravity multiplet with $n_T$
tensor multiplets, $n_V$ vector multiplets and $n_H$
hypermultiplets. Generic string and M-theory compactifications may
produce all of these multiplets with arbitrary values of $n_T$.
Anomaly cancellation using the Green-Schwarz mechanism, however,
restricts these numbers by the constraint
\cite{Randjbar-Daemi:1985wc} \be n_H = n_V + 273- 29 n_T. \ee
Starting from the tensor multiplets, we will restrict to the case
$n_T=1$, where there exists a covariant Lagrangian description of
the model. For this case, the constraint given above reduces to
 \be \label{e-2-2} n_H = n_V + 244. \ee
Regarding the hypermultiplets, the $4n_H$ hyperscalars must
parameterize a non-compact quaternionic manifold, whose possible
forms are given in \cite{Nishino:1986dc}; here, we will consider
the case where the hyperscalar manifold is $Sp(n_H,1) / Sp(n_H)
\times Sp(1)_R$. Finally, the vector multiplets must belong to a
gauge group that is a subgroup of the $Sp(n_H,1)$ isometry group
of the quaternionic manifold, possibly times extra factors under
which the hypermultiplets transform as singlets (the $E_6$ in
\cite{Randjbar-Daemi:1985wc} is one such example). Aside from such
factors, one usually chooses the gauge group to be a product of a
subgroup of $Sp(n_H)$ and a subgroup of $Sp(1)_R$. Under the first
factor, the hyperinos may transform in arbitrary representations
while the gravitino and tensorino are inert. Under the second
factor, the hyperinos are inert (although the hyperscalars are
charged) while the gravitino, tensorino and gauginos transform
non-trivially. Here, we will consider a $U(1)_R$ subgroup in which
case the gravitino, the tensorino and the gauginos as well as the
hyperscalars have all unit charge.

In our model, we pick the gauge group to be $E_7 \times G_2 \times
U(1)_R$. Thus, we have a total of $133+14+1=148$ vector multiplets
and Eq. (\ref{e-2-2}) requires that the total number of
hypermultiplets be equal to $148+244=392$. The hyperinos fit
nicely (no singlets!) into a half-hypermultiplet of the pseudoreal
784-dimensional representation $(\mathbf{56},\mathbf{14})$. So,
the transformation properties of the various fermions under the
three gauge group factors are as follows \bea \label{e-2-3}
\psi^{-}_M &:&\quad (\mathbf{1},\mathbf{1})_1, \nn\\
\chi^{+} &:&\quad (\mathbf{1},\mathbf{1})_1, \nn\\
\lambda^{-} &:&\quad (\mathbf{133},\mathbf{1})_1 +
(\mathbf{1},\mathbf{14})_1 +
(\mathbf{1},\mathbf{1})_1, \nn\\
\psi^{+} &:&\quad \frac{1}{2} (\mathbf{56},\mathbf{14})_0, \eea
where the subscripts indicate $U(1)_R$ charges.

Passing on to the hyperscalars, they parameterize the manifold \be
\label{e-2-4} \MM = \frac{Sp(392,1)}{Sp(392) \times Sp(1)_R}, \ee
where the holonomy group in the denominator corresponds to the
(unbroken) local symmetry of the scalar sector. One may then gauge
a subgroup of the isometry group $Sp(392,1)$; in our case we will
consider the gauging of $E_7 \times G_2 \times U(1)_R$. The
embedding of $E_7\times G_2$ in $Sp(392)$ is defined by
identifying $(\mathbf{56},\mathbf{14})$ with the pseudoreal
fundamental representation $\mathbf{784}$ of $Sp(392)$.

The construction of the gauged supergravity theory proceeds along
the steps described in \cite{Nishino:1986dc,Nishino:1997ff}; for
the $E_7 \times E_6 \times U(1)_R$ case, this procedure was
outlined in \cite{Randjbar-Daemi:1985wc} and discussed in great
detail in \cite{Randjbar-Daemi:2004qr} and can be applied with
minor modifications to our model as well. The only aspects of this
construction that need attention refer to the hyperscalar sector.
To begin, we let $\alpha=1,\ldots,4 \times 392$ label the
coordinates on $\MM$ and $a=1,\ldots,2 \times 392$ label the
fundamental of $Sp(392)$. We then pick a gauge-fixed coset
representative $L$ and we decompose its Maurer-Cartan form into
the coset vielbein and the $Sp(392)$ and $Sp(1)_R$ connections
\bea \label{e-2-5}
V_\alpha^{ai} &=& ( L^{-1} \partial_\alpha L )^{ai} ,\nn\\
\MA_\alpha^{ab} &=& ( L^{-1} \partial_\alpha L )^{ab} ,\nn\\
\MA_\alpha^{ij} &=& ( L^{-1} \partial_\alpha L )^{ij}, \eea whose
pullbacks on the spacetime manifold define the composite vielbein
and connections \bea \label{e-2-6}
P_M^{ai} &=& ( L^{-1} \partial_M L )^{ai} = \partial_M \varphi^\alpha V_\alpha^{ai} ,\nn\\
Q_M^{ab} &=& ( L^{-1} \partial_M L )^{ab} = \partial_M \varphi^\alpha \MA_\alpha^{ab} ,\nn\\
Q_M^{ij} &=& ( L^{-1} \partial_M L )^{ij} = \partial_M
\varphi^\alpha \MA_\alpha^{ij}, \eea that are used to construct
scalar kinetic terms and spinor covariant derivatives respectively.
Gauging $E_7 \times G_2 \times U(1)_R$ entails introducing the $E_7$
gauge fields $A_M^I$, $I=1,\ldots,133$, the $G_2$ gauge fields
$A_M^{I'}$, $I'=1,\ldots,14$ and the $U(1)_R$ gauge field $A_M^3$
and replacing ordinary derivatives by gauge-covariant ones. In
particular, the covariant derivative acting on the hyperscalars is
\be \label{e-2-7} \MD_M \varphi^\alpha = \partial_M \varphi^\alpha -
g ( A_M^I \xi^{I\alpha} + A_M^{I'} \xi^{I' \alpha} ) - g' A_M^3
\xi^{3\alpha}, \ee where $T^I$, $T^{I'}$ and $T^3$ are the
antihermitian $E_7$, $G_2$ and $U(1)_R$ generators, $\xi^{I\alpha} =
(T^I \varphi)^\alpha$, $\xi^{I'\alpha} = (T^{I'} \varphi)^\alpha$
and $\xi^{3\alpha} = (T^3 \varphi)^\alpha$ are Killing vectors
associated with the respective isometries and $g$ and $g'$ are the
$E_7 \times G_2$ and $U(1)_R$ couplings. Accordingly, the composite
vielbein and connections in (\ref{e-2-6}) are replaced by the gauged
versions \bea \label{e-2-8}
\MP_M^{ai} &=& ( L^{-1} \MD_M L )^{ai} = \MD_M \varphi^\alpha V_\alpha^{ai} ,\nn\\
\MQ_M^{ab} &=& ( L^{-1} \MD_M L )^{ab} = \MD_M \varphi^\alpha \MA_\alpha^{ab} ,\nn\\
\MQ_M^{ij} &=& ( L^{-1} \MD_M L )^{ij} = \MD_M \varphi^\alpha
\MA_\alpha^{ij} - g' A_M^3 (T^3)^{ij}. \eea

A direct consequence of the gauging of $U(1)_R$ is the emergence
of a scalar potential in the theory. This comes about due to the
fact that the commutator $[\mathcal{D}_M , \mathcal{D}_N]
\epsilon^i$ appearing in the supersymmetry variation of the
gravitino kinetic term acquires an extra term involving the gauge
field strengths and the functions \bea \label{e-2-9}
&C^I_{ij} = ( L^{-1} T^I L )_{ij} ,\quad C^{I'}_{ij} = ( L^{-1} T^{I'} L )_{ij} , \nn\\
&C^3_{ij} = ( L^{-1} T^3 L )_{ij}. \eea Restoring local
supersymmetry requires then a set of modifications to the Lagrangian
and transformation rules; in the bosonic sector, this induces a
hyperscalar potential.

The Lagrangian of the gauged supergravity theory was first derived
in \cite{Nishino:1986dc} and further elaborated upon in
\cite{Nishino:1997ff}. Its bosonic part may be written as
{\allowdisplaybreaks \bea \label{e-2-10}
e^{-1} \ML &=& \frac{1}{4} R - \frac{1}{4} \partial_M \phi \partial^M \phi - \frac{1}{12} e^{2 \phi} G_{MNP} G^{MNP} \nn\\
&& - \frac{1}{4} e^\phi v_A \tr( F_{A MN} F_A^{MN} )\nn\\
&& - g_{\alpha\beta} ( \varphi ) \MD_M \varphi^\alpha \MD^M \varphi^\beta \nn\\
&& + \frac{1}{8} e^{-1} \epsilon^{MNPQRS} B_{MN} v_A \tr( F_{A PQ} F_{A RS} ) \nn\\
&& - \frac{1}{4} e^{-\phi} [ g^2 ( v_7^{-1} C^I_{ij} C^{I ij} + v_2^{-1} C^{I'}_{ij} C^{I' ij} ) \nn\\
&& \qquad\qquad + g'^2 v_1^{-1} C^3_{ij} C^{3 ij} ]. \eea}Here,
the summation index $A=7,2,1$ runs over the three gauge group
factors, $v_\alpha$ are a set of constants to be determined later
and the various traces are interpreted as e.g. $\tr(F_{7 MN}
F_7^{MN}) = F^I_{MN} F^{I MN}$. In this form of the Lagrangian,
the field strength of $B_{MN}$ is given by the usual definition
$G_{MNP} = 3 \partial_{[M} B_{NP]}$ and the Green-Schwarz term is
explicit.

\section{Anomaly Cancellation}
\label{sec-3}

In this section, we demonstrate that the theory described above is
anomaly-free. Generally, $D=6$ chiral supergravities suffer from
gravitational, gauge and mixed anomalies arising from box diagrams
with four external gravitons and/or gauge bosons and one chiral
spinor or (anti-)self-dual 2--form running in the loop. Starting
from the gravitational anomalies, we first note that the
contributions from the 2--forms from the supergravity and tensor
multiplets cancel each other so that the only nonzero terms come
from the gravitino, tensorino, gauginos and hyperinos. Summing
these contributions (in that order) using the formulas of the
Appendix, we find the expression \bea \label{e-3-1}
I_8(R) &=& - \left[ \frac{49}{72} \tr R^4 - \frac{43}{288} (\tr R^2)^2 \right] \nn\\
&& + (1 - 148 + 392 ) \left[ \frac{1}{360} \tr R^4 + \frac{1}{288} (\tr R^2)^2 \right] \nn\\
&=& (\tr R^2)^2, \label{e-5-1-12} \eea which confirms that the
irreducible $\tr R^4$ terms cancel and explains the particular
normalization chosen. Turning to the gauge anomalies, these may be
split into (i) $E_7$ and $G_2$ anomalies (contributions from
gauginos and hyperinos), (ii) $U(1)_R$ anomalies (contributions from
gravitino, tensorino and gauginos), (iii) $E_7 \times G_2$ anomalies
(contributions from hyperinos) and (iv) $E_7 \times U(1)_R$ and $G_2
\times U(1)_R$ anomalies (contributions from gauginos). Writing down
the various contributions in the order indicated above, we find
{\allowdisplaybreaks \bea \label{e-3-2}
I_8(F) &=& \frac{2}{3} ( - \Tr F_7^4 + 7 \tr F_7^4  ) + \frac{2}{3} ( - \Tr F_2^4 + 28 \Tr F_2^4 ) \nn\\
&& + \frac{2}{3} ( -5 + 1 - 148 ) F_1^4 \nn\\
&& + 2 \tr F_7^2 \Tr F_2^2 \nn\\
&& - 4 \Tr F_7^2 F_1^2 - 4 \Tr F_2^2 F_1^2 \nn\\
&=& - \frac{2}{3} \Tr F_7^4 + \frac{14}{3} \tr F_7^4 + 18 \Tr F_2^4 - \frac{304}{3} F_1^4 \nn\\
&& + 2 \tr F_7^2 \Tr F_2^2 - 4 \Tr F_7^2 F_1^2 - 4 \Tr F_2^2 F_1^2.
\eea }where ``$\tr$'' and ``$\Tr$'' stand for fundamental and
adjoint traces respectively. Finally, we have to consider the mixed
anomalies. Splitting them into (i) mixed $E_7$ and $G_2$ anomalies
(contributions from gauginos and hyperinos) and (ii) mixed $U(1)_R$
anomalies (contributions from gravitino, tensorino and gauginos), we
write {\allowdisplaybreaks \bea \label{e-3-3}
I_8(F,R) &=& \frac{1}{6} \tr R^2 ( \Tr F_7^2 - 7 \tr F_7^2 + \Tr F_2^2 - 28 \Tr F_2^2 ) \nn\\
&& + \frac{1}{6} ( -19 - 1 + 148 ) \tr R^2 F_1^2 \nn\\
&=& \tr R^2 \left( \frac{1}{6} \Tr F_7^2 - \frac{7}{6} \tr F_7^2 - \frac{9}{2} \Tr F_2^2 \right) \nn\\
&& + \frac{64}{3} \tr R^2 F_1^2. \eea }Eqs.
(\ref{e-3-2}-\ref{e-3-3}) can be simplified by expressing all
traces in the fundamental representations. Moreover, since both
$E_7$ and $G_2$ factors do not possess fourth-order invariants,
all of the above traces can be expressed exclusively in terms of
second-order traces. For the two factors, we have the identities
\bea \label{e-3-4}
\Tr F_7^2 = 3 \tr F_7^2, &\qquad& \Tr F_2^2 = 4 \tr F_2^2, \nn\\
\tr F_7^4 = \frac{1}{24} (\tr F_7^2)^2, &\qquad& \tr F_2^4 = \frac{1}{4} (\tr F_2^2)^2, \nn\\
\Tr F_7^4 = \frac{1}{6} (\tr F_7^2)^2, &\qquad& \Tr F_2^4 =
\frac{5}{2} (\tr F_2^2)^2. \eea Substituting these in (\ref{e-3-2})
and (\ref{e-3-3}), we write the gauge and mixed anomalies in the
simplified form \bea \label{e-3-5}
\!\!\!\!\!\!\!\!I_8(F) &=& \frac{1}{12} (\tr F_7^2)^2 + 45 (\tr F_2^2)^2 - \frac{304}{3} F_1^4 \nn\\
&& + 8 \tr F_7^2 \tr F_2^2 - 12 \tr F_7^2 F_1^2 - 16 \tr F_2^2
F_1^2, \eea and \be \label{e-3-6} I_8(F,R) = \tr R^2 \left( -
\frac{2}{3}  \tr F_7^2 - 18 \tr F_2^2 + \frac{64}{3} F_1^2 \right).
\ee Putting the three contributions (\ref{e-3-1}), (\ref{e-3-5}) and
(\ref{e-3-6}) together, we find that the total anomaly polynomial is
given by {\allowdisplaybreaks \bea \label{e-3-7}
\!\!I_8 &=& (\tr R^2)^2 \nn\\
&& + \tr R^2 \left( - \frac{2}{3}  \tr F_7^2 - 18 \tr F_2^2 + \frac{64}{3} F_1^2 \right) \nn\\
&& + \frac{1}{12} (\tr F_7^2)^2 + 45 (\tr F_2^2)^2 - \frac{304}{3} F_1^4 \nn\\
&& + 8 \tr F_7^2 \tr F_2^2 - 12 \tr F_7^2 F_1^2 - 16 \tr F_2^2
F_1^2. \eea } In order for the Green-Schwarz mechanism to operate,
the above polynomial must factorize as \be \label{e-3-8} I_8 = (
\tr R^2 + u_A \tr F_A^2 ) ( \tr R^2 + \tilde{u}_A \tr F_A^2 ). \ee
To check whether this is possible, one must (i) compare the $(\tr
F_A^2)^2$ and $\tr R^2 \tr F_A^2$ terms to determine
$(u_A,\tilde{u}_A)$ and (ii) check if, for the values of
$(u_A,\tilde{u}_A)$ thus determined, the $\tr F_A^2 \tr F_B^2$
cross-terms match as well. Remarkably, it turns out that all
conditions are indeed satisfied with \bea \label{e-3-9}
u_7 = - \frac{1}{2}&,& \quad u_2 = - 3, \quad u_1 = -4, \nn\\
\tilde{u}_7 = - \frac{1}{6}&,& \quad \tilde{u}_2 = - 15, \quad
\tilde{u}_1 = \frac{76}{3}. \eea Cancellation of anomalies is then
a straightforward matter. One may set the undetermined constants
$v_A$ in (\ref{e-2-10}) equal to \be \label{e-3-10} v_A=-u_A, \ee
and modify the $B_{MN}$ gauge transformation law to \be
\label{e-3-11} \delta B_2 \sim \omega^1_{2L} + \tilde{u}_A
\omega^1_{2Y,A}, \ee where $\omega^1_{2L}$ and $\omega^1_{2Y,A}$
are related to $\tr R^2$ and $\tr F_A^2$ by descent. The
corresponding anomalous variation of the Green-Schwarz term in
(\ref{e-2-10}) (including a gravitational term) cancels exactly
the variation of the effective action.

Apart from the perturbative anomalies discussed above, our model
may also have global anomalies. In particular, since the $G_2$
factor in the gauge group has a non-trivial sixth homotopy group,
$\pi_6 (G_2) = \MBBZ_3$, there exist gauge transformations not
continuously connected to the identity. Under such
transformations, the effective action may pick up a phase factor
and is thus ill-defined. The condition for the absence of global
anomalies in the case of the $G_2$ gauge group is given by
\cite{Bershadsky:1997sb} \be \label{e-3-12} 1 - 4 \sum_{\MR}
n_{\MR} b_{\MR} = 0 \mod 3, \ee where $n_{\MR}$ is the number of
hypermultiplets transforming in the representation $\MR$ of $G_2$
and $b_{\MR}$ is defined by $\tr_{\MR} F_2^4 = b_{\MR} ( \tr F_2^2
)^2$. In our model, the hypermultiplets are in the adjoint of
$G_2$ and we have $n_{\mathbf{14}}=\frac{1}{2} \times 56 = 28$ and
$b_{\mathbf{14}}=\frac{5}{2}$. For these numbers, the condition
(\ref{e-3-12}) is indeed satisfied and thus the theory is free of
global anomalies as well.

\section{Compactification}
\label{sec-4}

The bosonic Lagrangian  (\ref{e-2-10}) contains a hyperscalar
potential given by its last term. Employing the definitions $g_7 = g
/ \sqrt{v_7}$, $g_2 = g / \sqrt{v_2}$ and $g_1 = g' / \sqrt{v_1}$,
we rewrite this potential as \be \label{e-4-1} V(\varphi) =
\frac{1}{4} e^{-\phi} ( g_7^2 C^I_{ij} C^{I ij} + g_2^2 C^{I'}_{ij}
C^{I' ij} + g_1^2 C^3_{ij} C^{3 ij} ), \ee where the various
$C$--functions are given in (\ref{e-2-9}). In
\cite{Randjbar-Daemi:2004qr}, it was shown that a convenient
parameterization of the scalar coset is given by the $784 \times 2$
matrix \be \label{e-4-2} \varphi = \left(
\begin{array}{c} \varphi_1 \\ \vdots \\ \varphi_{392} \end{array}
\right), \ee where $\varphi_n$, $n=1,\ldots,392$, are themselves $2
\times 2$ matrices satisfying the reality condition
$\varphi^*_n=\sigma_2 \varphi_n \sigma_2$. With this
parameterization, one may define the coset representative as the
$(784 + 2) \times (784 + 2)$ matrix \bea \label{e-4-3} L = \left(
\begin{array}{cc} 1 + \left( \frac{\sqrt{1 + \varphi^\dag \varphi}
-1}{\varphi^\dag \varphi} \right) \varphi \varphi^\dag & \varphi
\\ \varphi^\dag & \sqrt{1 + \varphi^\dag \varphi} \end{array}
\right), \eea where the factor inside parentheses is understood as a
scalar (since $\varphi^\dag \varphi$ is proportional to the
identity). Using the definition (\ref{e-4-3}), it can be shown that
the $C$--functions take the form \bea \label{e-4-4} &C^I_{ij} = (
\varphi^\dag T^I \varphi )_{ij} ,\quad C^{I'}_{ij} =
( \varphi^\dag T^{I'} \varphi )_{ij} , \nn\\
&C^3_{ij} = \left[ 1 + \tr (\varphi^\dag \varphi) \right]
(T^3)_{ij}. \eea Then the potential is given by the simple
expression \bea \label{e-4-5} V(\varphi) &=& \frac{1}{16} e^{-\phi}
\left[ - g_7^2
( \varphi^\dag T^I \varphi )^2 - g_2^2 ( \varphi^\dag T^{I'} \varphi )^2 \right] \nn\\
&& + \frac{1}{8} e^{-\phi} g_1^2 \left[ 1 + \tr (\varphi^\dag
\varphi) \right]^2. \eea Using this expression and recalling that
$T^I$ and $T^{I'}$ are antihermitian, we immediately see that the
potential is strictly positive-definite and attains its unique
global minimum at $\varphi^\alpha = 0$. Among other things, this
implies that the $E_7 \times G_2$ gauge symmetry cannot be
spontaneously broken by the hyperscalars at tree level.

At $\varphi^\alpha =0$, the potential (\ref{e-4-5}) takes the
exponential form \be \label{e-4-6} V_{\min} = \frac{1}{8}
e^{-\phi} g_1^2, \ee and, for the case of constant $\phi$, it acts
like a cosmological constant. It is this effective cosmological
constant and the particular form of the dilaton coupling which
picks up the $\MBBR^4\times \MBFS^2$ among other
maximally-symmetric spaces \cite{Salam:1984cj}. In the
non-supersymmetric theory, de Sitter or anti- de Sitter spaces
would also be possible solutions \cite{Randjbar-Daemi:1982hi}.

In the $E_7 \times E_6 \times U(1)_R$ model of
\cite{Randjbar-Daemi:1985wc}, the monopole was embedded in the
$E_6$ factor, yielding a $D=4$ effective theory with $SO(10)
\times U(1)_R$ gauge symmetry where all massless fermions
originate from the $E_6$ gauginos since only the latter couple to
the monopole; in \cite{Randjbar-Daemi:2004qr} this was generalized
to include monopole embeddings in $E_7$ where zero-mode fermions
arise from the hyperinos as well. In the model considered here,
one can embed as many magnetic monopoles as the rank of the gauge
group which is 10. In general we will thus have a maximum of 10
monopole charges. This will give rise to fermionic zero modes from
the associated gauginos but, since the hyperinos are charged under
both $E_7$ and $G_2$ gauge group factors, it will also necessarily
give rise to fermionic zero modes from the associated hyperinos.
In the absence of a \emph{vev} for the vector potential associated
to $U(1)_R$, the gravitino, tensorino and the rest of the gauginos
will be massive. Turning to the bosons, the squared mass of each
one of the lightest hyperscalar fluctuations will receive two
contributions, one being proportional to the associated eigenvalue
of $\frac{\partial^2 V}{\partial \varphi^\alpha \partial
\varphi^\beta}$ at $\varphi^\alpha=0$ and the other being
proportional to $D^2$ where $D$ is the covariant derivative acting
on the hyperscalar fluctuations in the background of the monopole
vector potential(s). The first contribution will make all
hyperscalars massive. The second contribution, if the monopole
charges do not add up to zero, will be a positive quantity
proportional to $1/a^2$, where $a$ is the radius of $\MBFS^2$.
Furthermore in the case of a nonzero net monopole charge of the
hyperscalar even the leading (lightest) $D=4$ scalar modes
resulting from it will belong to a non trivial irreducible
representation of the Kaluza-Klein $SU(2)$. We shall comment on
the masses of some other bosonic modes in Section \ref{sec-6}.

In the absence of a \emph{vev} for the $U(1)_R$ gauge field, the
supersymmetric variation of the gravitino will be nonzero and thus
this class of compactifications will break all supersymmetries.

To write down the ansatz for monopole compactification, we employ
the rescalings $A^I_M\to A^I_M /g$, $A^{I'}_M\to A^{I'}_M /g$ and
$A^3_M \to A^3_M / g'$ and we set the metric, the $U(1) \subset
E_7$ gauge field and the dilaton equal to \bea \label{e-4-7}
&ds_6^2 = \eta_{\mu\nu} dx^\mu dx^\nu + a^2 ( d \theta^2 + \sin^2 \theta d \varphi^2 ), \nn\\
&A_\pm = \frac{n}{2} Q ( \cos \theta \mp 1 ) d \varphi; \quad F = \frac{n}{2} Q \sin \theta d \theta\wedge d \varphi, \nn\\
&\phi = \phi_0 = \textrm{const.} \eea Here, $A_+$ and $A_-$
correspond to the potentials on the northern and southern
hemisphere which, on the equator, should be connected by a gauge
transformation parameterized by $U=e^{ i n Q \varphi}$. In order
for $U$ to be single-valued as $\varphi$ changes by $2 \pi$, the
quantity $n q_{\min}$, where $q_{\min}$ is the minimal $U(1)_M$
charge in the theory, must be an integer. In the above, $Q$ is any
generator of the gauge group and, in general, can be a linear
combination of all commuting generators with appropriate
quantization conditions on the coefficients. This ansatz solves
the field equations \cite{Randjbar-Daemi:1985wc}.

To prepare the setting for the spectrum analysis of the next
section let us consider some examples. A first example is given by
embedding $U(1)_M$ in $E_7$ according to the maximal-subgroup
decomposition \be \label{e-4-12} E_7 \supset E_6 \times U(1). \ee
Using the branching rules \cite{Slansky:1981yr}\bea \label{e-4-13}
\mathbf{56} &\to& \mathbf{27}_1 + \mathbf{\overline{27}}_{-1} + \mathbf{1}_{3} + \mathbf{1}_{-3}, \nn\\
\mathbf{133} &\to& \mathbf{78}_0 + \mathbf{27}_{-2} +
\mathbf{\overline{27}}_{2} + \mathbf{1}_{0}, \eea we see that
$q_{\min} = 1$ so that $n = \mathrm{integer}$. Discarding neutral
fields, we see that the fermion representations under $E_6 \times
G_2 \times U(1)_M$ which can give rise to fermion zero modes on
$\MBFS^2$ are, \be \label{e-4-14} (\mathbf{27},\mathbf{14})_1 +
(\mathbf{\overline{27}},\mathbf{14})_{-1} +
(\mathbf{1},\mathbf{14})_{3} + (\mathbf{1},\mathbf{14})_{-3}, \ee
for the $E_7$ hyperinos and \be \label{e-4-15}
(\mathbf{27},\mathbf{14})_{-2} +
(\mathbf{\overline{27}},\mathbf{14})_{2}, \ee for the $E_7$
gauginos.

As a second example, consider the successive maximal-subgroup
decompositions \be \label{e-4-16} E_7 \supset SO(12) \times SU(2)
\supset SO(10) \times SU(2) \times U(1), \ee and identify the last
$U(1)$ factor with $U(1)_M$. Using the branching rules
 \bea \label{e-4-17}
\mathbf{56} &\to& (\mathbf{12},\mathbf{2}) + (\mathbf{32},\mathbf{1}), \nn\\
\mathbf{133} &\to& (\mathbf{1},\mathbf{3}) +
(\mathbf{32'},\mathbf{2}) + (\mathbf{66},\mathbf{1}), \eea for $E_7
\supset SO(12) \times SU(2)$ and \bea \label{e-4-18}
\mathbf{12}  &\to& \mathbf{1}_1 + \mathbf{1}_{-1} + \mathbf{10}_0, \nn\\
\mathbf{32}  &\to& \mathbf{16}_1 + \overline{\mathbf{16}}_{-1}, \nn\\
\mathbf{32'}\!\! &\to& \mathbf{16}_{-1} + \overline{\mathbf{16}}_1, \nn\\
\mathbf{66}  &\to& \mathbf{1}_0 + \mathbf{10}_2 + \mathbf{10}_{-2}
+ \mathbf{45}_{0}, \eea for $SO(12) \supset SO(10) \times U(1)$,
we see again that, $q_{\min} = 1$ and $n = \mathrm{integer}$.
Discarding neutral fields, we find that the muliplets of $SO(10)
\times SU(2) \times G_2 \times U(1)_M$ which can have fermion zero
modes on $\MBFS^2$ are\bea \label{e-4-19}
&&(\mathbf{1},\mathbf{2},\mathbf{14})_1 + (\mathbf{1},\mathbf{2},\mathbf{14})_{-1} \nn\\
&&+ (\mathbf{16},\mathbf{1},\mathbf{14})_1 +
(\mathbf{\overline{16}},\mathbf{1},\mathbf{14})_{-1}, \eea for the
hyperinos and \bea \label{e-4-20}
&&(\mathbf{16},\mathbf{2},\mathbf{14})_{-1} + (\overline{\mathbf{16}},\mathbf{2},\mathbf{14})_1 \nn\\
&&+ (\mathbf{10},\mathbf{1},\mathbf{14})_{2} +
(\mathbf{10},\mathbf{1},\mathbf{14})_{-2}, \eea for the $E_7$
gauginos. In the above examples, the representations
(\ref{e-4-14}) and (\ref{e-4-19}) are understood as
half-hypermultiplets.

\section{The Chiral Spectrum and (In-)stabilities}
\label{sec-5}

The monopole embeddings discussed in the previous section give
rise to many chiral fermions in the complex representations of the
unbroken gauge group in the effective $D=4$ theory. Using the
formalism of \cite{Randjbar-Daemi:1982hi} we can evaluate the
content of the effective $D=4$ theory. Here we apply this
formalism to the two examples of the previous section.

In the first example the unbroken gauge group is $E_6\times
G_2\times U(1)_R\times SU(2)_{KK}$, where $SU(2)_{KK}$ denotes the
Kaluza-Klein gauge group originating from the isometries of
$\MBFS^2$. The chiral fermions originate from the $\mathbf{27}$'s
and the $\mathbf{\overline{27}}$'s. We can regard all the $D=4$
fermions as left-handed Weyl spinors. The chiral fermions
originating from the decomposition of $\mathbf{56}$ of $E_7$ then
are \be 2 (\mathbf{\overline{27}}, \mathbf{14}, \mathbf{n})_{0} ,
\ee while the fermions originating from the decomposition of the
adjoint of $E_7$ produce \be (\mathbf{27} , \mathbf{14},
\mathbf{2n})_{1} + (\mathbf{27}, \mathbf{14}, \mathbf{2n})_{-1},
\ee with the subscripts here denoting the $U(1)_R$ charges.

In the second example the unbroken gauge group in $D=4$ is $G=
SO(10) \times SU(2)\times G_2\times SU(2)_{KK}\times U(1)_R$. The
spectrum of the $D=4$ chiral fermions is given by \be
2(\mathbf{\overline {16}}, \mathbf{1}, \mathbf{14},\mathbf{n})_0,
\ee and \be (\mathbf{16},\mathbf{2},\mathbf{14},\mathbf{n})_1 +
(\mathbf{16}, \mathbf{2}, \mathbf{14}, \mathbf{n})_{-1}. \ee

It is clearly seen that the spectrum in both cases is free from
all chiral anomalies, because $E_6$ and $SO(10)$ are safe groups
in $D=4$ and the $U(1)_R$ couplings are obviously vectorlike. It
is also seen that there is no value of $n$ which produces a
realistic spectrum. One can study other embeddings with the aim of
reducing the gauge group and the number of families. For example,
the group $G_2$ can be broken completely by the embedding of a
monopole in an $SU(2)$ subgroup of $G_2$ relative to which the
branching is $\mathbf{14} = \mathbf{3} + \mathbf{11}$. By itself
this will produce only a vectorlike theory in $D=4$ with an
unbroken group $E_7 \times SU(2)_{KK}\times U(1)_R$. However,
combined with other monopoles in the manner described above, one
can break the group down to $SO(10)\times SU(2)_{KK}$. The number
of families will nevertheless be still large.

Apart from the proliferation of the number of families and other
shortcomings for a realistic  model building with tree level
considerations (such as the absence of a realistic Higgs spectrum
and Yukawa couplings at tree level), there is a fundamental
difficulty with all such compactifications. In fact it seems that,
with the exception of the simplest compactifying solution in which
the monopole is embedded in the $U(1)_R$ factor, all other
compactifications are unstable. To see this, let us embed the
momopole in one or both of the non-Abelian factors. Denote by $V$
one of the excitations of the vector potential tangent to
$\MBFS^2$ and charged with respect to $U(1)_M$. This vector has
the components $V_{\pm}$ with respect to a complex basis in the
tangent space of $\MBFS^2$. We also have the reality condition $V=
V^{\dagger}$. As a Lie-algebra-valued vector we can write $V$ as
$V = U_{+} ^{r} T_r + W_{+} ^ {r}T_r^{\dag}$, where $U$ and $W$
are complex fields and the $T$'s are among the charged generators
of the gauge group. For example they can be the generators of
$E_7$ in the directions of $\mathbf{10}_2$ or $\mathbf{16}_1$ of
the previous section. In order to be able to write down a general
formula which can be applied for any model of this kind, denote
the $U(1)_M$ charge of $U$ or $W$ by $q$. The mass spectrum of
$D=4$ spin-zero fields resulting from such a $D=6$ object is given
by \cite{Randjbar-Daemi:1983bw} \be a^2 M^2 = \ell (\ell+1) -
(\lambda -1)^2, \ee where $\ell = |\lambda|, |\lambda|+1, ...$ and
$\lambda =  1 + \frac{n}{2} q$. It is easy to see that for all
those fields for which $ nq \leq -2$ there is a tachyon. For
example, with positive $n$, the leading mode in the spectrum of
$\mathbf{10}_2$ will be a tachyon of squared mass $-n/2a^2$.

The only way to avoid this conclusion is to find an embedding for
which $|nq|=1$ for all the excitations. With integer $n$ and $q$
we then need to have $n= \pm 1$ and $ q=\pm 1$ for all the fields.
Such an embedding is guaranteed to exist in all cases where there
is a gauge group factor $G_A$ that has a maximal-subgroup
decomposition $G_\alpha \supset H \times U(1)$ with $H$ simple and
where all fermions charged under this group transform in the
adjoint. This is exactly what happens in the $E_7 \times E_6\times
U(1)_R$ model, where the monopole can be embedded in the
``hidden'' $E_6$ that gives rise to $\mathbf{16}$'s of $SO(10)$
with $q =\pm 1$. The monopole with the minimal charge of $\pm 1$
gives thus a stable compactification with two chiral families of
$SO(10)$ in the 16-dimensional representation transforming as
singlets under the unbroken $E_7\times SU(2)_{KK}$ but charged
relative to the unbroken $U(1)_R$. It seems difficult to obtain an
analogous result for the new model presented in this paper
essentially because any embedding which produces integer $q$'s
necessarily contains fields for which $|q| \geq 2$. The only
exception is of course the half-supersymmetric solution in which
the monopole is embedded in $U(1)_R$. This embedding will leave
$E_7\times G_2$ unbroken.

\section{Conclusions}
\label{sec-6}

In this paper, we have demonstrated the existence of a second
consistent $\MN=1$ gauged supergravity model in six dimensions
besides the old $E_7 \times E_6 \times U(1)_R$ model. The model
found here is based on the $E_7 \times G_2 \times U(1)_R$ gauge
group, with hypermatter transforming as a half-hypermultiplet in
the pseudoreal representation $(\mathbf{56},\mathbf{14})$. The
theory satisfies a set of very stringent anomaly constraints and
turns out to be free of local and global anomalies. Also, as its
$E_7 \times E_6 \times U(1)_R$ sibling or, in fact, any other
model of this type, the theory admits a monopole compactification
on $\MBBR^4 \times \MBFS^2$. Despite the fact that embedding
monopole type configurations in $E_7\times G_2$ will produce many
chiral fermions in $D=4$, it seems that all such solutions are
perturbatively unstable. On the other hand, if the monopole is
identified with the vector potential of $U(1)_R$ only, a stable
compactification will be obtained and, with the choice of $n=1$,
half of the $D=6$ supersymmetries will be unbroken. The model will
clearly inherit all the brane solutions discovered so far in the
context of $\MN=1$ supergravity models in six dimensions, to some
of which we referred in Section \ref{sec-1}.

As far as the tachyons are concerned, one may adopt the point of
view that they are welcome in the context of an effective theory
as they are natural candidates for $D=4$ Higgs fields. The quartic
term in the potential for such fields will come from the
self-coupling of the $D=6$ gauge fields and their \emph{vev} will
break the $E_6\times SU(2)_{KK}$ (or $SO(10)\times SU(2)_{KK}$) at
the KK scale. Such an origin for the Higgs fields has been
considered before as a possible solution to the hierarchy problem.
In order for this interpretation to be complete, one needs to look
for new stable solutions of the six-dimensional field equations
which would correspond to the minimum of the potential for the
tachyons interpreted as Higgs fields. These solutions will
necessarily break the spherical symmetry and their construction
may give a geometrical origin to the Higgs mechanism. It will be
interesting to find such solutions.

On the other hand, for phenomenological reasons, we may want to
prevent certain modes from becoming tachyonic. In this paper we
gave a necessary and sufficient condition for this to happen.
Namely, in order for an excitation of an internal component of the
gauge field not to be tachyonic, it is sufficient that $|nq|=1$,
where $nq$ is understood as the sum of the individual $nq$'s over
all the monopole directions with respect to which the
corresponding excitation is charged.

A final question, motivated by the uniqueness of the $D=6$ gauged
supergravities under consideration and the fact that they cannot
be constructed through straightforward string or M-theory
compactifications, refers to their origin in terms of a
higher-dimensional fundamental theory. Although previous
experience might suggest that such models possibly arise due to
some new mechanism involving non-perturbative physics, such a
mechanism has not been identified up to date.

\vskip .2in

\appendix

\section{Anomaly Polynomials}

The anomaly structure of $D=6$ theories is encoded in a set of
formal eight-forms, called anomaly polynomials. In our
conventions, the gravitational anomaly polynomials
\cite{Alvarez-Gaume:1983ig} are given by {\allowdisplaybreaks \bea
\label{e-a-1}
I^{1/2}_{8} (R) &=& \frac{1}{360} \tr R^4 + \frac{1}{288} (\tr R^2)^2, \nn\\
I^{3/2}_{8} (R) &=& \frac{49}{72} \tr R^4 - \frac{43}{288} (\tr R^2)^2, \nn\\
I^A_{8} (R) &=& \frac{7}{90} \tr R^4 - \frac{1}{36} (\tr R^2)^2.
\eea}The gauge anomaly polynomials are given by {\allowdisplaybreaks
\bea \label{e-a-2}
I^{1/2}_{8} (F) &=& \frac{2}{3} \tr F^4, \nn\\
I^{1/2}_{8} (F_A,F_B) &=& 4 \tr F_A^2 \tr F_B^2, \nn\\
I^{3/2}_{8} (F) &=& \frac{10}{3} \tr F^4, \eea}where the second
polynomial applies to the case where product representations are
present. Finally, the polynomials corresponding to mixed anomalies
are {\allowdisplaybreaks \bea \label{e-a-3}
I^{1/2}_{8} (F,R) &=& - \frac{1}{6} \tr R^2 \tr F^2, \nn\\
I^{3/2}_{8} (F,R) &=& \frac{19}{6} \tr R^2 \tr F^2. \eea}Here, the
superscripts $1/2$, $3/2$ and $A$ refer to a spin $1/2$ fermion, a
spin $3/2$ fermion and a 2--form potential respectively. The above
anomaly polynomials correspond to Weyl spinors of positive chirality
and 2--form potentials with self-dual field strengths. For
negative-chirality spinors or anti-self-dual field strengths, the
sign of the anomaly is reversed.

\vskip .2in

\noindent {\bf Acknowledgement.} This work is supported by the
EPEAEK programmes ``Heraclitus" and ``Pythagoras" and co-funded by
the European Union (75\%) and the Hellenic State (25\%).

\end{document}